\documentclass[aps,prl,twocolumn,groupedaddress]{revtex4}
\usepackage{amsmath,amssymb}

\usepackage{graphicx}

\usepackage{hyperref}

\newenvironment{itemize*}%
  {\setlength{\parskip}{-0.0em}
   \begin{itemize}%
    \setlength{\itemsep}{-0.4em}%
    }%
  {\end{itemize}\vskip-0.8em}
  
\newcommand{\xsection}[1]{\vspace{-0.5em}\section{#1}\vspace{-0.5em}}
\newcommand{\xsubsection}[1]{\vspace{-1em}\subsection{#1}\vspace{-1em}}

\def\longer#1{}
\def\short#1{#1}

\def\dt{\frac{\partial}{\partial t}}
\def\Psitx{\Psi(t,\vec{x})}
\def\psitx#1{\psi_{#1}(t,\vec{x})}
\def\psix#1{\psi_{#1}(\vec{x})}
\def\H{{\cal H}}
\def\R{\mathbb{R}}
\def\ra{\rightarrow}
\def\la{\lambda}
\renewcommand{\Re}{\mbox{Re}}
\renewcommand{\Im}{\mbox{Im}}

\def\refa#1{(\ref{#1})}
\def\mybox#1{{\fbox{#1}}}

\begin{document}

\title{Spontaneous particle creation in time-dependent overcritical fields}

\author{Nikodem Szpak}
\email[e-mail: ]{szpak@th.physik.uni-frankfurt.de}
\affiliation{Institut f{\"u}r Theoretische Physik, J.W.Goethe Universit{\"a}t, 60054 Frankfurt/Main, Germany}
\affiliation{Albert-Einstein-Institut, MPI f{\"u}r Gravitationsphysik, Am M{\"u}hlenberg 1, 14476 Golm, Germany}
\date{\today}

\begin{abstract}
It is believed that in presence of some strong electromagnetic fields, called \textit{overcritical}, the (Dirac) vacuum becomes unstable and decays leading to a spontaneous production of an electron-positron pair. Yet, the arguments are based mainly on the analysis of static fields and are insufficient to explain this phenomenon completely. Therefore, we consider time-dependent overcritical fields and show, within the external field formulation, how spontaneous particle creation can be defined and measured in physical processes. We prove that the effect exists always when a \textit{strongly overcritical} field is switched on, but it becomes unstable and hence generically only approximate and non-unique when the field is switched on and off. In the latter case, it becomes unique and stable only in the adiabatic limit.
\end{abstract}


\maketitle

\xsection{Introduction}

In a long debate \cite{MuRaGr72b, RaMuGr74, ReGr77, KlaSch77b, SchSei82, Nen80, Nen87, Pickl-PhD} on whether and how spontaneous particle creation (of $e^+e^-$ pairs) can be uniquely defined as an effect of the vacuum decay in presence of overcritical (electromagnetic) fields either static or adiabatic fields have been considered, what does not really answer the question in a realistic physical situation. The main problem in time-dependent overcritical fields is to distinguish between two sources of particle creation: dynamical, due to the time-dependence, and spontaneous, due to the overcriticality of the external field. In \cite{NS-PhD} we have studied analytically and numerically various time-dependent overcritical fields and shown when the effect can be uniquely defined and behaves stable. The aim of this Letter is to summarize our results which will appear in a a more extensive form elsewhere.

\xsubsection{Classical Dirac equation}

Electrons in an external time-dependent electromagnetic field $A_\mu(x)$ are described by the Dirac equation which can be written in the evolution form
\begin{equation} \label{DiracH} \nonumber
i\dt\Psitx = H(t) \Psitx \quad\mbox{with}\quad H(t)\equiv H_0+V(t),
\end{equation}
where $V(t) = eA_0+e\alpha^i A_i$ is the time-dependent external potential and
$H_0 = -i \hbar c\alpha^i \partial_i + m c^2 \beta$ is the free Hamiltonian\longer{\footnote{For a large class of potentials the Hamiltonians $H(t)$ are essentially self-adjoint on ${\cal C}^\infty_0(\R^3)$ and can be uniquely extended to self-adjoint operators $\hat{H}(t)$ on ${\cal L}^2(\R^n)$. For brevity we identify $H$ with $\hat{H}$.}}.
Consider every $H(t)$ separately treating $t$ as a parameter. For atomic-like (localized) potentials the spectrum $\sigma(H)$ has two continuous parts $(-\infty,-mc^2)\cup(mc^2,\infty)$ and a possible discrete part $\left\{E_n\in(-mc^2,mc^2)\right\}$. The corresponding wave functions $\psix{E}$, satisfying $H\psi_{E} = E\psi_{E}$, describe: electron scattering states for $E>mc^2$, bound states for $|E|<mc^2$ and positron scattering states for $E<-mc^2$.
\longer{
 \begin{center}
 \includegraphics[width=0.7\linewidth]{spectrum}
 \end{center}
}
\xsubsection{{Overcriticality on the classical level}}

\begin{figure}[h]
  \includegraphics[width=0.7\linewidth]{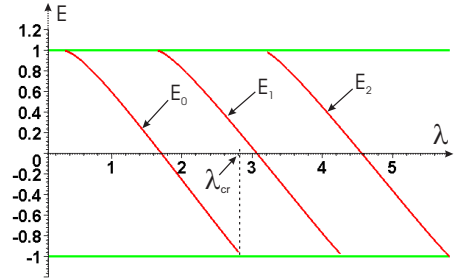}
  \caption{Spectrum of $H_\la=H_0+\la\,V$ as a function of $\la$.\label{fig-boundstates}}
\end{figure}

Consider a one-parameter family of potentials $V_\la(\vec{x}) \equiv \la \cdot V(\vec{x})$.
Enumerate the bound-state energies: $E_0(\la)<E_1(\la)<E_2(\la)<...$. 
Then the bound-state energies $E_n(\la)$ depend continuously on $\la$.
Assume $V(\vec{x})$ is negative (attractive for electrons). 
Then if $\la$ increases, $E_n(\la)$ decrease towards $-mc^2$ (fig. \ref{fig-boundstates}). There exists a large class of potentials $V(\vec{x})$ such that $E_0(\la_{cr})=-mc^2$ for a finite $\la=\la_{cr}$, called crititcal.
For $\la>\la_{cr}$ the bound-state $E_0$ disappears from the spectrum. 
Such potentials $V_\la$ are called \textit{overcritical}.
Further, next bound states $E_1$, $E_2, ...$, disappear as $\la$ grows.

\xsubsection{{Resonances}}

\begin{figure}[h] 
  \mybox{\includegraphics[width=0.6\linewidth]{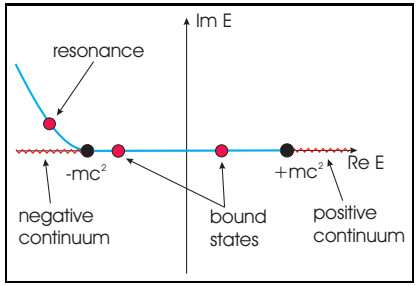}} 
  \caption{Position of the real or complex pole of the resolvent $R_\la$ meaning a bound state or a resonance, respectively.\label{resonance}}
\end{figure}

Bound-states crossing the value $E=-mc^2$ are forbidden to get embedded in the negative continuum, so they disappear from the spectrum and turn into resonances, which can be traced as poles (on the analytic continuation) of the resolvent $R_\la\equiv (H_\la-E)^{-1}$ (fig. \ref{resonance}).

As the Hamiltonian changes so that a bound-state turns into a resonance, the dominant part of the bound state wave function forms a wave packet localized spectrally in the continuum around the \textit{resonance energy} 
$E_R \equiv \Re\,E$. The half-width of the packet is equal to $\Gamma\equiv \Im\,E$. During evolution generated by a static overcritical Hamiltonian the wave packet decays spatially (but stays localized spectrally).

\xsubsection{What is expected by overcriticality ?}

\begin{figure}[h]
  \includegraphics[width=0.8\linewidth]{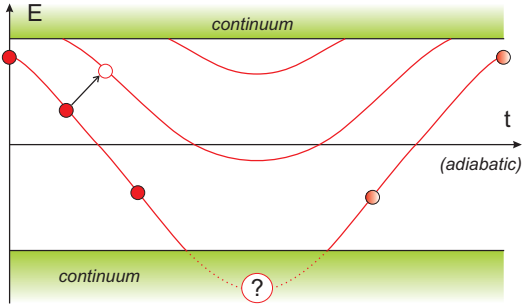}
  \caption{Spectrum $\sigma(H(t))$ of time-dependent Hamiltonian.\label{E_t}}
\end{figure}

Consider time-dependent Hamiltonians $H(t)=H_0+\la(t)\,V$ with an overcritical period $\la(t)>\la_{cr}$ for $|t|<T$ (fig. \ref{E_t}).
When $\la(t)$ changes adiabatically the bound states (defined at every instant $t$) change slowly and according to the adiabatic theorem the wave function ``follows'' them. Yet, there is no adiabatic theorem for resonances, thus during the overcritical period the wave packet may decay and stay trapped in the continuum forever.
\begin{figure}[h]
  \includegraphics[width=0.8\linewidth]{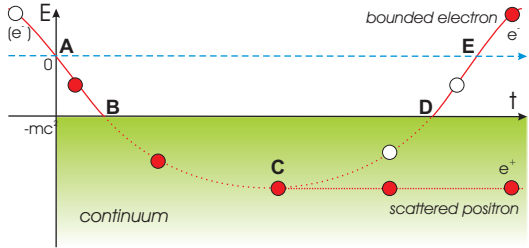}
  \caption{Pair production in an overcritical process.\label{E_t_overcr_phases}}
\end{figure}

To explain these processes in terms of particle creation and annihilation one needs a many--particle description, i.e. the second quantized Dirac theory. Formulated in a non-rigorous language it provides the following scenario \cite{NS-PhD} (see fig. \ref{E_t_overcr_phases}): \longer{Increasing the strength of the potential, an empty particle bound-state crosses the boundary between particle and antiparticle subspaces and turns into an occupied antiparticle bound-state, what we call \textit{weak overcriticality}. Increasing the strength of the potential further, the bound-state turns into a resonance and the wave function forms a wave packet in the negative continuum, what we call \textit{strong overcriticality}. Then, keeping the potential constant, the wave packet decays. Next, decreasing the strength of the potential back to a weakly overcritical, an empty antiparticle bound-state reappears while the whole wave function stays trapped in the negative continuum. Going back to a subcritical value of the potential, the empty antiparticle bound-state turns into an occupied particle bound-state.}\short{A) An empty particle bound-state crosses the boundary between particle and antiparticle subspaces and turns into an occupied antiparticle bound-state, what we call \textit{weak overcriticality}. B) The bound-state turns into a resonance and the wave function forms a wave packet in the negative continuum, what we call \textit{strong overcriticality}. C) The wave packet decays. D) An empty antiparticle bound-state reappears while the whole wave function stays trapped in the negative continuum. E) The empty antiparticle bound-state turns into an occupied particle bound-state.} Finally, we obtain a pair: a bound particle and a scattered antiparticle.
\begin{figure}[h] 
  \includegraphics[width=0.8\linewidth]{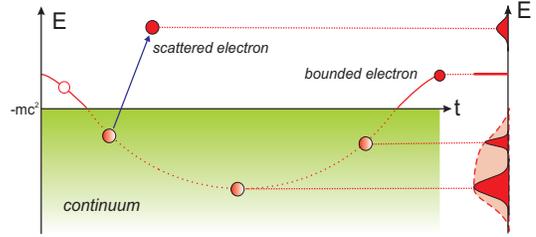} 
  \caption{Spectrum of produced particles and antiparticles in result of a time-dependent overcritical field.}
\end{figure}

In practice, except the adiabatic case, the wave function disperses during evolution over the whole spectrum what results in an additional \textit{dynamical} pair creation. Essential is the question whether it is possible to separate the \textit{spontaneous particle creation} occuring due to the overcriticality of the potential from the dynamical one.

\def\H{{\cal H}}
\def\F{{\cal F}}
\def\vx{\vec{x}}
\def\psitx#1#2{\psi_{#1}^{(#2)}(t,\vec{x})}
\newcommand{\phix}[2][]{\phi_{#2}^{#1}(\vec{x})}
\newcommand{\phit}[2][]{\phi_{#2}^{#1}}
\newcommand{\PsiF}[2][]{\hat{\Psi}_{#1}(#2)}
\def\PsiS#1{\hat{\Psi}^*(#1)}
\renewcommand{\b}[2][]{\hat{b}_{#1}(#2)}
\newcommand{\bS}[2][]{\hat{b}_{#1}^*(#2)}
\renewcommand{\d}[2][]{\hat{d}_{#1}(#2)}
\newcommand{\dS}[2][]{\hat{d}_{#1}^*(#2)}
\newcommand{\bn}[2][]{\hat{b}^{#1}_{#2}}
\newcommand{\bnS}[2][]{\hat{b}^{#1*}_{#2}}
\newcommand{\dn}[2][]{\hat{d}^{#1}_{#2}}
\newcommand{\dnS}[2][]{\hat{d}^{#1*}_{#2}}
\renewcommand{\Psitx}[1][]{\hat{\Psi}_{#1}(t,\vec{x})}
\def\Spm{S_{+-}}
\def\Smp{S_{-+}}
\def\dt{\frac{\partial}{\partial t}}

\def\Om{\Omega}
\def\hU{\hat{U}}
\def\hV{\hat{V}}
\def\hQ{\hat{Q}}
\def\hH{\hat{H}}
\def\hS{\hat{S}}

\xsection{QED in external fields}

To describe physical processes one needs full QED, but it appears too difficult to be solved in our case, so we consider the external field approximation, i.e. treat the electromagnetic field as classical and quantize only the Dirac field. This approximation is believed to be accurate as long as the number of charged Dirac particles stays small and does not influence the electromagnetic field.

In presence of overcritical fields the vacuum and particles change their properties qualitatively so that in order to treat them correctly the analysis must start from the first principles.

\xsubsection{{Fock space and the field operator}}

The construction of the theory begins with the algebra of fields $\PsiF{f}$ anti-linear in $f\in\H$ -- the canonical anticommutation relations
\begin{eqnarray} \nonumber
  \{\PsiF{f},\PsiS{g}\}_+ &=& (f,g)\\
  \{\PsiF{f},\PsiF{g}\}_+ &=& \{\PsiS{f},\PsiS{g}\}_+ = 0 \nonumber
\end{eqnarray}
$\forall f,g\in\H$ with $(f,g)$ a scalar product in $\H$. We represent them as self-adjoint operators acting in the Fock space. The representation is unique up to the choice of a pair of projectors $P_\pm: \H\rightarrow \H_\pm$ which split the Hilbert space $\H \equiv \H_+ \bigoplus \H_-$ on the particle and antiparticle subspaces. Then, the Fock space is constructed as 
\begin{equation} \nonumber
  \F =   \bigoplus_{m,n=0}^\infty \F^{(n,m)}
  = \bigoplus_{m,n=0}^\infty 
  \overbrace{\underbrace{\H_+ \otimes ... \otimes \H_+}_{\mbox{n times}}
     \otimes \underbrace{\H_- \otimes ... \otimes \H_-}_{\mbox{m times}}}^{\mbox{antisymmetric}}
\end{equation}
with $\F^{(0,0)}$ one-dimensional space with a unit vector $\Omega$, a no-particle state called \textit{vacuum}.
Using the projectors $P_\pm$ we split
\begin{equation} \nonumber
  \PsiF{f} = \b{P_+ f} + \dS{P_- f}\qquad \forall f\in\H
\end{equation}
so that $\b{P_+ f}\,\Om = \d{P_- f}\,\Om = 0$. Then $\bS{f}$ and $\dS{g}$ create particles and antiparticles in states $f\in\H_+$ and $g\in\H_-$, respectively, and $\b{f}$ and $\d{g}$ annihilate them. 

The choice of the projectors $P_\pm$ is equivalent to the choice of the vacuum vector $\Om\in\F$. Two representations based on $P_\pm$ and $P'_\pm$ are unitary equivalent if and only if the Hilbert-Schmidt norm
\begin{equation} \label{P-P}
  ||P_\pm - P'_\pm||_{H.S.} < \infty.
\end{equation}
So unitarily non-equivalent representations, giving different physical predictions, are possible. To exclude them one distinguishes $P_\pm$ as spectral projections on the positive and negative energy subspaces of the Hamiltonian $H$. Then the induced vacuum vector $\Om$ is a ground state of the Hamiltonian $\hH$ implemented in the Fock space $\F$.

\xsubsection{{Particle scattering in Fock space}}

For time-dependent Hamiltonians $H(t)$ the above construction must be repeated at every instant of time with time-dependent projectors $P_\pm(t)$ and leads to a family of Fock spaces $\F(t)$. To implement scattering processes it is sufficient to consider only the initial $\F_{in}$ and final $\F_{out}$. The classical (one-particle) unitary scattering operator $S:\H\ra\H$ is implemented in the Fock space by a unitary $\hat{S}: \F \ra \F$ such that
\begin{equation} \nonumber
  \PsiF[out]{f} \equiv \PsiF[in]{S f} = \hat{S} \;\PsiF[in]{f}\; \hat{S}^*.
\end{equation}
It exists when $S$ satisfies the Shale-Stinespring criterion
\begin{equation} \nonumber
 ||S_{\pm\mp}||_{H.S.} \equiv ||P_\pm S P_\mp||_{H.S.} < \infty
\end{equation}
This guarantees that the initial vacuum state $\Omega\in\F$ evolves to $\hat{S}\;\Omega\in\F$.

\xsection{Structure of $\hS_\la$}

For time-dependent processes with $H(t)\equiv H_0 + V(t)$ the scattering operator $\hS$ has the following form
\begin{equation} \label{S-S0-S1}
  \hS = C_0\; :\hS_0\ \widetilde{S}: 
\end{equation}
\longer{where
  \begin{equation}
    C_0 \equiv [\det(1+A^* A)]^{-1/2},
  \end{equation}
  \begin{equation} \label{S1}
  \begin{split}
    \widetilde{S}^* = :
     &\exp\left(\sum_{k,l} A_{kl} \bnS{k} \dnS{l}\right) 
     \exp\left(\sum_{k,l} (B_{kl}-\delta_{kl})\bnS{k} \bn{l}\right) \\
     &\exp\left(\sum_{k,l} (C_{kl}-\delta_{kl})\dnS{k} \dn{l}\right)
     \exp\left(\sum_{k,l} D_{kl} \bn{k} \dn{l}\right): \\
     &k=n_++1,...,\infty;\; l=n_-+1,...,\infty
  \end{split}
  \end{equation}
  \begin{equation} \label{S0}
  \begin{split}
    \hS^*_0\equiv\;:&\left(\bnS{1}\mp\sum_{k=1}^{n_+} (S_{-+})_{k1}\dn{k}\right)
                   ...\left(\bnS{n_+}\mp\sum_{k=1}^{n_+} (S_{-+})_{kn_+}\dn{k}\right)\\
                  &\left(\dnS{1}\mp\sum_{k=1}^{n_-} \overline{(S_{+-})_{k1}}\bn{k}\right)
               ...\left(\dnS{n_-}\mp\sum_{k=1}^{n_-}\overline{(S_{+-})_{kn_-}}\bn{k}\right):
  \end{split}
  \end{equation}
  \begin{equation*}
    n_+= \dim\ker S_{++}, \qquad n_-= \dim\ker S_{--}
  \end{equation*}
and
  \begin{align} \nonumber
    A &\equiv -S^{-1}_{++}\;S_{+-},& B&\equiv \pm S^{-1}_{++}, \\
    C &\equiv \pm\overline{(S^{-1}_{--})},& D&\equiv \pm(S_{-+}\;S^{-1}_{++})^T. \label{S-ABCD}
  \end{align}
}  
\short{where $C_0$ is a normalization constant, $\widetilde{S}$ creates and annihilates particle--antiparticle pairs or scatters, and the exceptional part $\hS_0$ creates and annihilates single particles.}
Acting on vacuum
\begin{equation} \nonumber
  \hS\; \Om = C_0\; {\dnS{n_-} \dots \dnS{1}\; \bnS{n_+} \dots \bnS{1}}\;
  \exp\Big(\sum_{k,l} A_{kl} \bnS{k} \dnS{l}\Big) \Om
\end{equation}
$\hS_0$ creates single particles and antiparticles $\bnS{i}$ and $\dnS{i}$ which correspond to the \textit{special states} $\phi_i^\pm$, which are mapped by $S$ from $\H_\pm \equiv P_\pm(-\infty)\H$ to $\H'_\mp \equiv P'_\mp(+\infty)\H$. 
It has been conjectured that the spontaneous particle creation is associated with the presence of the exceptional part $\hS_0$ \cite{KlaSch77b}.

\xsubsection{{Potential switched on:} $H(-\infty)=H_0\neq H(+\infty)$}

For unequal initial and final Hamiltonians: $P_\pm \equiv P_\pm(-\infty) \neq P_\pm(+\infty) \equiv P'_\pm$, $\Om\neq \Om'_\la$ and $\bn{n}\neq\bn{n}{}'$, $\dn{n}\neq\dn{n}{}'$.
For a family of time-dependent processes $H_\la(t)\equiv H_0 + \la V(t)$ one can get $\hS(\la)$ according to \refa{S-S0-S1}\longer{-\refa{S-ABCD}}, but now we want to describe the final state in terms of the final vacuum $\Om'$ and final operators $\bn{n}{}',\dn{n}{}'$. To obtain this, we decompose the scattering into two parts: $\hS(\la) \;\Om \equiv \hV(\la)^*\;\hU(\la)\;\Om$, such that $\hU(\la)$ implements the scattering in $\F_{in}$, i.e. in terms of $\bn{n}{},\dn{n}{}$ and $\hV(\la)^*$ implements the change of projectors $\hV(\la)^*\;\Om = \Om'_\la$, i.e. the transition from $\F_{in}$ to $\F_{out}$ \cite[Th. 3]{NS-PhD}. We show that $\hU(\la)$ is analytic in $\la$ and hence has no stable \textit{special states} (w.r.t. $\la$). On the other hand, $\hV(\la)$ has stable \textit{special states} if and only if the final Hamiltonian $H(\infty)$ is overcritical! Then (for $n_-=1$ and $n_+=0$)
\begin{equation} \nonumber
  \hS(\la) \;\Om = C_0\; {\dnS{1}{}'} \exp\Big(\sum_{k,l} A_{kl} \bnS{k}{}' \dnS{l}{}'\Big) {\Om'}
\end{equation}
We conclude that creation of the antiparticle is \textit{not} due to the scattering process $\hU(\la)$ but due to the \textit{change in the structure of the vacuum}, which can be schematically expressed as $\Om \cong \dnS{1}{}'\; \Om'$.

The charge of the final state with respect to the initial vacuum and particles is conserved $Q=(\hS\;\Om, \hQ\;\hS\;\Om) = 0$, but with respect to the final vacuum and particles 
$Q'=(\hS\;\Om, \hQ'\;\hS\;\Om) = 1$ and reflects the existence of the an antiparticle. Simultaneously, the charge of the final vacuum with respect to the initial one is $(\Om', \hQ\;\Om') = -1$, hence we say the final vacuum is (relatively) charged.

Unfortunately, presently no processes with a switch-on of an overcritical potential are accessible experimentally.

\xsubsection{{Potential switched on and off:} $H(-\infty)=H(+\infty)=H_0$}

Analogously, time-dependent processes $H_\la(t)\equiv H_0 + \la V(t)$ give $\hS(\la)$. Since \cite[Th. 16]{NS-PhD} $\hS(\la)$ is analytic in $\la$, all special states are \textit{unstable} w.r.t. $\la$!

There are two ways to treat this problem. Either to defend the role of \textit{special states} in the definition of spontaneous particle creation and consider the adiabatic limit which is free from the above instability, or to relax the condition of \textit{special states} and define the spontaneous pair creation in a {\it weaker sense}.

\xsection{{Definition by the adiabatic limit}}

In the adiabatic limit the dynamical pair production tends to 0, while the spontaneous pair production survives the limit. Let's consider processes where the potential $V_{\la,\epsilon}(t,\vec{x}) \equiv \la\, e^{-\epsilon^2 t^2}\, \tilde{V}(\vec{x})$ varies arbitrarily slow in time ($\epsilon \ra 0$) and vanishes as $t\rightarrow\pm\infty$. We can calculate the scattering operators $\hat{S}_{\la,\epsilon}$. In the adiabatic limit it is possible that $\lim_{\epsilon\rightarrow 0} \hat{S}_{\la,\epsilon} \neq \hat{S}_{\la,0}$ and the probability of particle creation: $r_\la = 1 - \lim_{\epsilon\rightarrow 0} \left|\left( \Omega, \hat{S}_{\la,\epsilon} \Omega\right)\right|^2$ has a jump at $\la=\la_{cr}$, because\vspace{-0.8em}
\begin{itemize*} 
  \item  for $0<\la<\la_{cr}$ ($V_\la$ subcritical):\quad $r_\la \stackrel{\epsilon\rightarrow 0}{\longrightarrow} 0$,
  \item for $\la>\la_{cr}$ ($V_\la$ overcritical):\quad $r_\la \stackrel{\epsilon\rightarrow 0}{\longrightarrow}r_0>0$
\end{itemize*}
This conjecture, posed and proved in the subcritical case by Nenciu \cite{Nen80}, in the overcritical case has been recently proved by Pickl \cite{Pickl-PhD}. Moreover, it has been shown that the decay of the resonance occurs already at the edge of the negative continuum leading to a creation of an antiparticle with vanishing momentum.
\begin{figure}[h] 
  \includegraphics[width=0.8\linewidth]{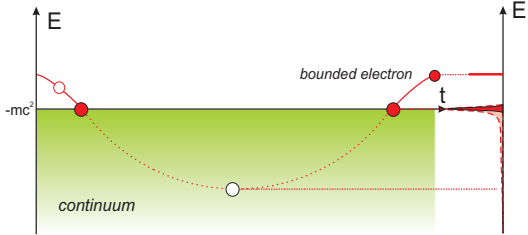} 
  \caption{Spectrum of produced particles in the adiabatic limit.}
\end{figure}

\xsubsection{Results: switch on and off processes}

\begin{itemize*} 
\item In all subcritical processes, when $E_0(t)>-mc^2$ exists for all $t$, no particles are created: $\hS\;\Om = \Om$.
\item In the case, when in some interval of time $t_1<t<t_2$ the bound state turns into a resonance in the negative continuum $\Re\, E_0(t)<-mc^2$, which we call \textit{strongly overcritical}, there is exactly one pair created spontaneously: $\hS\;\Om = \bS{\chi}\;\dS{\psi}\;\Om$.
\end{itemize*}

\xsubsection{Results: switch on processes}

\begin{itemize*}
\item When the final Hamiltonian is subcritical, the initial vacuum evolves to the final vacuum: $\hS\;\Om = \Om'$.
\item In the weakly overcritical case, i.e. $-mc^2<E_0(t)<0$ for all $t>t_0$, one bound antiparticle is created: $\hS\;\Om = \dS{\psi_B}{}'\;\Om'$.
\item In the strongly overcritical case, i.e. $\Re\,E_0(t)<-mc^2$ for all $t>t_0$, one free antiparticle is created: $\hS\;\Om = \dS{\psi_C}{}'\;\Om'$.
\end{itemize*}
The essential difference between the results in weakly and strongly overcritical processes is that $\psi_B$ is a localized bound state while $\psi_C$ is a dispersing wave packet in the continuum. Therefore, $\dS{\psi_C}{}'\;\Om'$ ``decays'', i.e. the antiparticle escapes to infinity and can be measured by a distant detector, while $\dS{\psi_B}{}'\;\Om'$ is a bounded neutral composite whose components cannot be detected in any experiment! Otherwise, it would violate the unitary equivalence of representations guaranteed by \refa{P-P} for two choices of the spectral projectors, namely the one used here, with $P_+$ projecting on $[0,+\infty)$, with another one (used by Greiner \textit{at al}, see \cite{GrMuRa} and references therein) where $\widetilde{P}_+$ projects on $(-mc^2,+\infty)$. In the latter case our weak overcriticality is indistinguishable from the subcritical situation. 

The same argument cannot be used to identify our strong overcriticality with a subcritical situation in some other representation, because then the projector $\widetilde{P}_+$ would necessarily include a part of the negative continuum, violate the condition \refa{P-P} and  lead to a unitarily nonequivalent representation.

We conclude that \textit{only the strong overcriticality leads to physically observable effects.}

\section{Spontaneous pair creation in a "weaker sense"}

Instability of the special states of $\hS(\la)$ means that
\begin{equation} \nonumber
  \hS(\la)\; \Om = C_0\; {\dS{\psi} \; \bS{\chi}}
  \exp\Big(\sum_{k,l} A_{kl} \bnS{k} \dnS{l}\Big) \Om,
\end{equation}
after perturbation $\la\ra\la+\delta\la$ goes over into
\begin{equation} \nonumber
  \hat{\tilde{S}}(\la)\; \Om = C_0\; \exp\Big(\sum_{k,l} \tilde{A}_{kl} \bnS{k} \dnS{l} +
  { B \dS{\tilde{\psi}} \; \bS{\tilde{\chi}}}\Big) \Om,
\end{equation}
with $|B|<1$, but by continuity $|B|\approx 1$. One can try to relax the condition of creation of a pair in the special states (with probability 1) in the definition of spontaneous particle creation to creation of a pair with probability $\approx 1$ in the corresponding states. However, it is difficult to keep this definition unique.


\bibliography{qed}
\bibliographystyle{unsrt}

\end{document}